\documentclass[prl,amsmath,amsfonts,amssymb,letterpaper,superscriptaddress,twocolumn,showpacs]{revtex4}
\pdfoutput=1
\usepackage{bm}
\usepackage{graphicx}
\usepackage{hyperref}
\usepackage{subfigure}

\hypersetup{
colorlinks=true,
    citecolor=blue,
    pdfborder={0 0 0},
}

\newcommand\arrow\vec

\def\vec#1{\bm{#1}}

\def\negspace{\!}

\def\lrsub#1#2#3{{\vphantom{#1}}_{#2} \negspace {#1} \negspace {\vphantom{#1}}_{#3}}

\def\bra#1{\left\langle {#1} \right\rvert}
\def\ket#1{\left\lvert {#1} \right\rangle}

\def\inprod#1#2{\left\langle {#1} | {#2} \right\rangle}

\def\inprodsubsub#1#2#3#4{\lrsub {\inprod{#1}{#2}} {#3} {#4}}

\def\pqinprod#1#2{\inprodsubsub{#1}{#2} p q}
\def\qpinprod#1#2{\inprodsubsub{#1}{#2} q p}
\def\pinprod#1#2{\inprodsubsub{#1}{#2} p p}
\def\qinprod#1#2{\inprodsubsub{#1}{#2} q q}
\def\pqbraket\pqinprod
\def\qpbraket\qpinprod
\def\pbraket\pinprod
\def\qbraket\qinprod

\def\outprod#1#2{\ket {#1}\!\bra {#2}}

\def\1{I}

\def\v0{{\bvec 0}}

\begin{document}

\title{Near-deterministic creation of universal cluster states with probabilistic Bell measurements and $3$-qubit resource states}

\author{Hussain A. Zaidi}
\email{haz4z@virginia.edu}
\affiliation{Department of Physics, University of Virginia, 382 McCormick Road, Charlottesville, Virginia 22904}
\affiliation{Department of Biochemistry and Molecular Biology, University of Virginia, 1340 Jefferson Park Avenue, Charlottesville, Virginia 22908, USA}

\author{Chris Dawson}
\affiliation{Quantitative Brokers, 2 Queen Caroline Street, London, W6 9DX, UK}

\author{Peter van Loock}
\affiliation{Institute of Physics, Johannes-Gutenberg Universit\"at Mainz, Staudingerweg 7, 55128 Mainz, Germany}

\author{Terry Rudolph}
\email{rudolph@imperial.ac.uk}
\affiliation{Controlled Quantum Dynamics Theory Group, Imperial College London, Prince Consort Road, London SW7 2BW, UK}


\date\today

\begin{abstract}
We develop a scheme for generating a universal qubit cluster state using probabilistic Bell measurements without the need for feed-forward or long-time quantum memories. Borrowing ideas from percolation theory we numerically show that using unambiguous Bell measurements that succeed with $75$\% success probability, one could build a cluster state with an underlying pyrochlore geometry such that the probability of having a spanning cluster in a chosen direction approaches unity in the limit of an infinite lattice size. The initial resources required for the generation of a universal state in our protocol are $3$-qubit cluster states that are within experimental reach and are a minimal resource for a Bell-measurement-based percolation proposal. Since single and multi-photon losses can be detected in Bell measurements, our protocol raises the prospect of a fully error-robust scheme. 
\end{abstract}

\maketitle

\emph{Introduction---} A realistic blueprint for a quantum computer will have to contend with non-deterministic quantum operations that succeed only probabilistically. It has been shown that one could do efficient quantum information processing using only linear-optical elements, feed-forward, and photon number resolving detectors (PNRDs), where the measurements subject to feed-forward induce the required non-linear transformations probabilistically \cite{knill2001}. The ideas of this linear-optical quantum computation proposal were later applied to the case of one-way quantum computation \cite{raussendorf2001} by using probabilistic linear-optical quantum operations along with feed-forward to create universal quantum resource states \cite{nielsen2004, browne2005} on which quantum algorithms could be implemented by suitably choosing ``easy" local measurements on the resource qubits. A major technological challenge in the above proposals was the requirement of feed-forward and, as a result, long-time quantum memories for creating the required cluster state. 

An important step in devising proposals  without the above-mentioned experimental challenges was presented in \cite{kieling2007, kieling2008}. These proposals used the ideas of percolation theory to show that one could create, without feedforward or long-time memories, a percolated lattice with some underlying geometry, which could then be renormalized to a universal cluster state, e.g. one with the geometry of a square lattice. Specifically, in Ref. \cite{kieling2008}, a proposal was put forward to create a pyrochlore lattice starting with 4-qubit cluster states using Type-I fusion gate \cite{browne2005} that succeeded with $50$\% success probability and consumed one qubit upon application. The downside of using a Type-I fusion gate is that this gate is not robust to photon losses, making the application of the ideas of \cite{kieling2007, kieling2008} practically demanding. 

On another note, Bell measurements form an important part of quantum computation and communication protocols \cite{mattle1996, duan2001, gottesman1999, knill2001}. It is known that without feed-forward, ancillary photons, or non-linear quantum operations, the success probability of a Bell measurement is bounded by $50$\% \cite{calsamiglia2001}. Recently, a few proposals came out \cite{grice2011, zaidi2013, ewert2014} that increased the success probability of unambiguous Bell measurements past $50$\% without feedforward by using Bell pairs, squeezers, or single photons, along with linear optics and PNRDs. Ref. \cite{grice2011} describes a scheme for unambiguous Bell measurement that has a success probability of $75$\% with ancillae Bell pairs. Ref \cite{ewert2014} uses single photon ancilla for $75$\% success probability, and Ref \cite{zaidi2013} uses squeezing for a success probability of more than $62$\% without any ancilla modes. These schemes allow for the detection of either single photon or multi-photon losses by counting the number of detected photons, which makes them appealing from the perspective of error correction. 

While the use of Bell measurements in entanglement distribution has been studied in the past \cite{duan2001, bose1998}, we deal with the specific question of using probabilistic unambiguous Bell measurements to create large quantum resource states that suffice for universal quantum computation, with the specific focus of minimizing the size of the initial resources needed. We propose creating a percolated pyrochlore lattice starting with $3$-qubit cluster states. We assume that these initial resource states along with either Bell pairs or single photons (depending on the chosen Bell measurement scheme) are available on demand. Since Bell measurements consume two qubits upon application, $3$-qubit initial states are a minimal resource for generating large cluster states. The fact that all the mentioned proposals for Bell measurements are robust to photon losses for either single or multiple photons means that experimental photon losses can be detected and potentially remedied. Other schemes that use $3$-qubit resources or single photons for generating universal states in an error-robust way without relying on ideas of percolation theory \cite{varnava2007, varnava2008} require considerable feed-forward and quantum memories, while percolation-based protocols are ballistic. The lack of feed-forward makes these protocols well-suited for integrated photonics architectures where the primary source of loss is expected to be photonic switching \cite{matthews2009, berry2009}. Background details on percolation theory in the context of generating universal resource states can be found in \cite{kieling2007, kieling2008}. 

We begin by describing the use of probabilistic Bell measurements to ``fuse" cluster states (in other words, to generate entanglement between two unentangled cluster states). We then show how $3$-qubit resource states can be 
fused together to form a large universal quantum state using Bell measurements that succeed with $75$\% probability. Specifically, we choose the underlying geometry of our cluster state to be that of a pyrochlore lattice with 
random deletions of bonds (entanglement) and vertices (qubits) due to the probabilistic nature of Bell measurements. Next, we provide numerical evidence that we are above the percolation threshold for our pyrochlore lattice, 
i.e. for $75\%$-success probability Bell measurements, the probability of having a spanning cluster in one direction approaches unity as the size of the lattice approaches infinity. This spanning cluster is a general graph (not 
a linear cluster state) and as shown in Table [\ref{tab:Scaling}], the ratio of qubits in a spanning cluster to the total number of qubits in the lattice goes up from $0.25$ to $0.34$ as the size of the lattice is increased. 
Finally, we look at the scaling behavior of the scheme as a function of the size of the lattice in three dimensions. That is, we estimate the required lattice size in the transverse directions such that the probability of having 
a spanning cluster in the lateral direction is above $90\%$. Note that either of the schemes for Bell measurements \cite{grice2011, ewert2014} would work in our scheme since they use resource states that are smaller than $3$ 
qubits. We would like to point out the concurrent work \cite{segovia2014}, where a different protocol with the same aims as ours is presented.  

\begin{figure}[htb]
\begin{center}
\begin{tabular}{c}
\includegraphics[width= .8 \columnwidth]{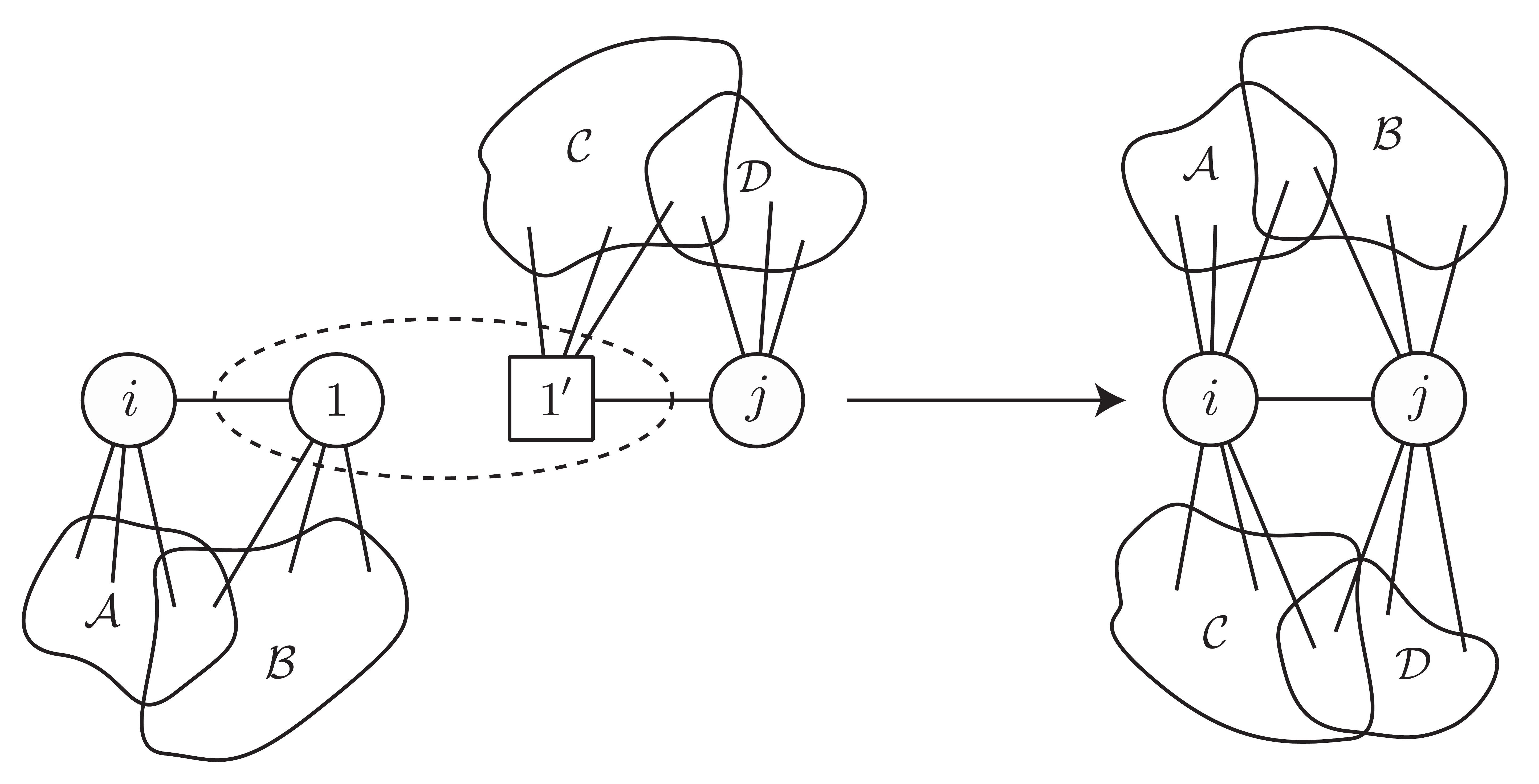}
\end{tabular}
\end{center}
\caption{The effect of a successful Bell measurement (measuring qubits $1$ and $1^\prime$) on qubits $i$ and $j$. A aquare node represents a polarization qubit with a Hadamard applied to it, while a dashed oblong around two qubits denotes a Bell measurement. The arguments presented in the appendix apply to all neighbours of the measured qubits, and hence, other neighbours inherit entanglement just as is shown for the qubits $i$ and $j$ in the figure. The italicized labels $\mathcal{A}$, $\mathcal{B}$, $\mathcal{C}$, and $\mathcal{D}$ denote graphs for which the products of Pauli $Z$ operators are represented as A, B, C, and D, respectively, in the appendix.}
\label{fig:BellMeasurementResult}
\end{figure}

\emph{Bell Measurements for Fusion---} Consider two cluster states that we would want to fuse into one as presented in Fig. [\ref{fig:BellMeasurementResult}]. Qubit $1^\prime$ has a Hadamard rotation applied to it before measurement. $\mathcal{A}$, $\mathcal{B}$, $\mathcal{C}$, and $\mathcal{D}$ represent the neighbours of qubits $i$, $1$, $1^\prime$, and $j$, respectively. For any chosen neighbours ($i$ and $j$) of measured qubits $1$ and $1^\prime$, Fig. [\ref{fig:BellMeasurementResult}] depicts the effect on $i$ and $j$ for a successful (i.e., unambiguous) Bell measurement.

The analysis that leads to Fig. [\ref{fig:BellMeasurementResult}] is similar to that presented in \cite{kok2010}, so we present the explicit steps in the appendix only. The main difference between our case and that presented in \cite{kok2010} is that we apply a Hadamard rotation to one of the qubits before measurement. The application of a Hadamard allows us to create cluster-like bonds between the neighbours of the measured qubits (as shown in the figure and the appendix), which turns out to be important for creating a lattice with a pyrochlore underlying geometry.  

An important question is what happens when a Bell measurement fails, that is, when we get an ambiguous output that could have come from two or more Bell inputs. Two arbitrary cluster states can always be represented by $\ket{A}=\ket{0}\ket{\xi}+\ket{1}\ket{\xi^\prime}$, and $\ket{B}=\ket{0}\ket{\Xi}+\ket{1}\ket{\Xi^\prime}$, where the primes denote that the neighbours of the first qubit have a Pauli $Z$ operation applied to them. Applying a Hadamard on the first qubit of $\ket{B}$, and writing the target qubits in the Bell basis yields
\begin{align}
\ket{A}\ket{B} &\sim  \left(\ket{\phi^+}+\ket{\phi^-}+\ket{\psi^+}-\ket{\psi^-}\right)\ket{\xi}\ket{\Xi}\nonumber\\ 
&- \left(\ket{\phi^+}+\ket{\phi^-}-\ket{\psi^+}+\ket{\psi^-}\right)\ket{\xi}\ket{\Xi^\prime}\nonumber\\
&+\left(\ket{\phi^+}-\ket{\phi^-}+\ket{\psi^+}+\ket{\psi^-}\right)\ket{\xi^\prime}\ket{\Xi} \nonumber\\
&+ \left(\ket{\phi^+}-\ket{\phi^-}-\ket{\psi^+}-\ket{\psi^-}\right)\ket{\xi^\prime}\ket{\Xi^\prime} 
\end{align}

Notice that when the Bell measurement results in an unambiguous output, we get $\ket{\xi} ( \ket{\Xi} + \ket{\Xi^\prime} ) + \ket{\xi^\prime} ( \ket{\Xi} - \ket{\Xi^\prime} )$ as the conditional state for the remaining qubits (up to local unitaries), which is indeed entangled between the qubits of $\ket{A}$ and those of $\ket{B}$. To analyze the failure mechanism, say that our Bell measurement fails to distinguish between $\ket{\phi^+}$ and $\ket{\phi^-}$, which is a probabilistic failure event in \cite{grice2011, ewert2014}. Further, assume that the magnitude of the coefficient of the ambiguous output is the same for both the inputs, which is also the case for \cite{grice2011, ewert2014}. Then the output looks like $\ket{\xi}(\ket{\Xi}-\ket{\Xi^\prime})$ or $\ket{\xi^\prime}(\ket{\Xi}-\ket{\Xi^\prime})$. This failure behavior means different consequences for different input graph states. The two case that are relevant for us are that of a 3-qubit triangle cluster input and a 3-qubit chain cluster input. 

\begin{figure}[htb]
\begin{center}
\begin{tabular}{c}
\includegraphics[width= 0.5 \columnwidth]{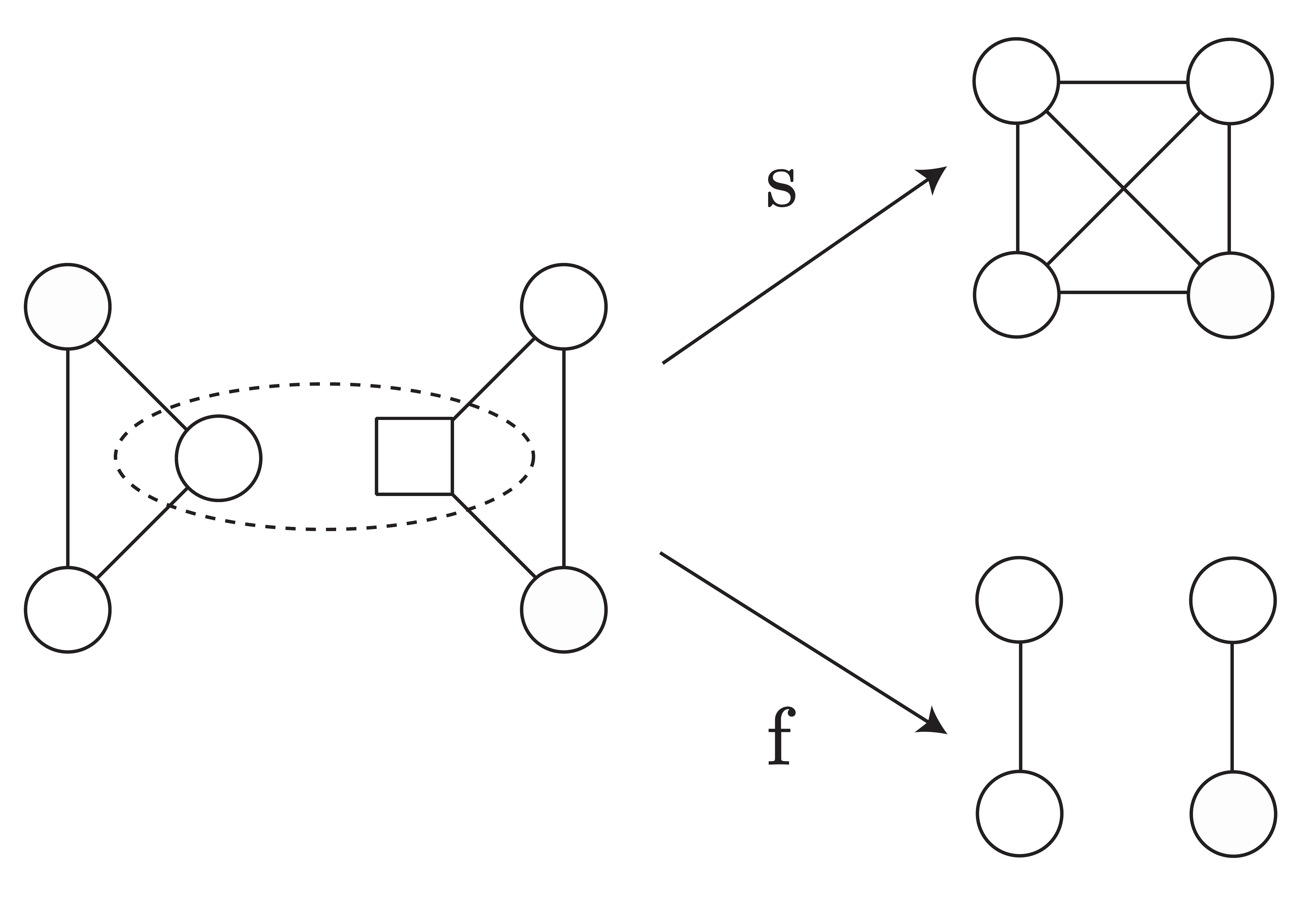}
\end{tabular}
\end{center}
\caption{The first step is to fuse two triangular clusters with Bell measurements. Upon success (denoted by ``s") we get a $4$-qubit tetrahedron. Failure of the measurement (denoted by ``f", and implying an ambiguous measurement outcome) results in two Bell pairs (up to local operations), which can still be used in subsequent fusions shown later.}
\label{fig:TriangleFusion}
\end{figure}

A 3-qubit triangular cluster is given by $\sim \ket{0}(\ket{0+}+\ket{1-}) +\ket{1}(\ket{0-}-\ket{1+})$. Hence, upon failing to fuse two triangular clusters we are left with a pair of Bell states up to local rotations, as shown in Fig. [\ref{fig:TriangleFusion}]. On the other hand, if we are trying to fuse a $3$-qubit cluster chain, $\sim \ket{0}(\ket{0+}+\ket{1-}) +\ket{1}(\ket{0+}-\ket{1-})$ (with Hadamard applied to one qubit), with another arbitrary state, then upon failure we are left with just $\ket{1-}$ for the qubits of the chain, i.e. the qubits of the chain are disentangled from each other. Notice that in both the above cases, the outcome for the cluster on which we do not apply a Hadamard is just that the measured qubit is destroyed and the rest of the graph stays intact up to local unitary operations. 

\emph{Pyrochlore Lattice Scheme---} The above success and failure mechanisms suffice to construct a percolated pyrochlore lattice. First, we fuse $3$-qubit triangular clusters to form tetrahedra as shown in Fig. [\ref{fig:TriangleFusion}].

\begin{figure}[htb]
\begin{center}
\begin{tabular}{c}
\includegraphics[width= 0.7 \columnwidth]{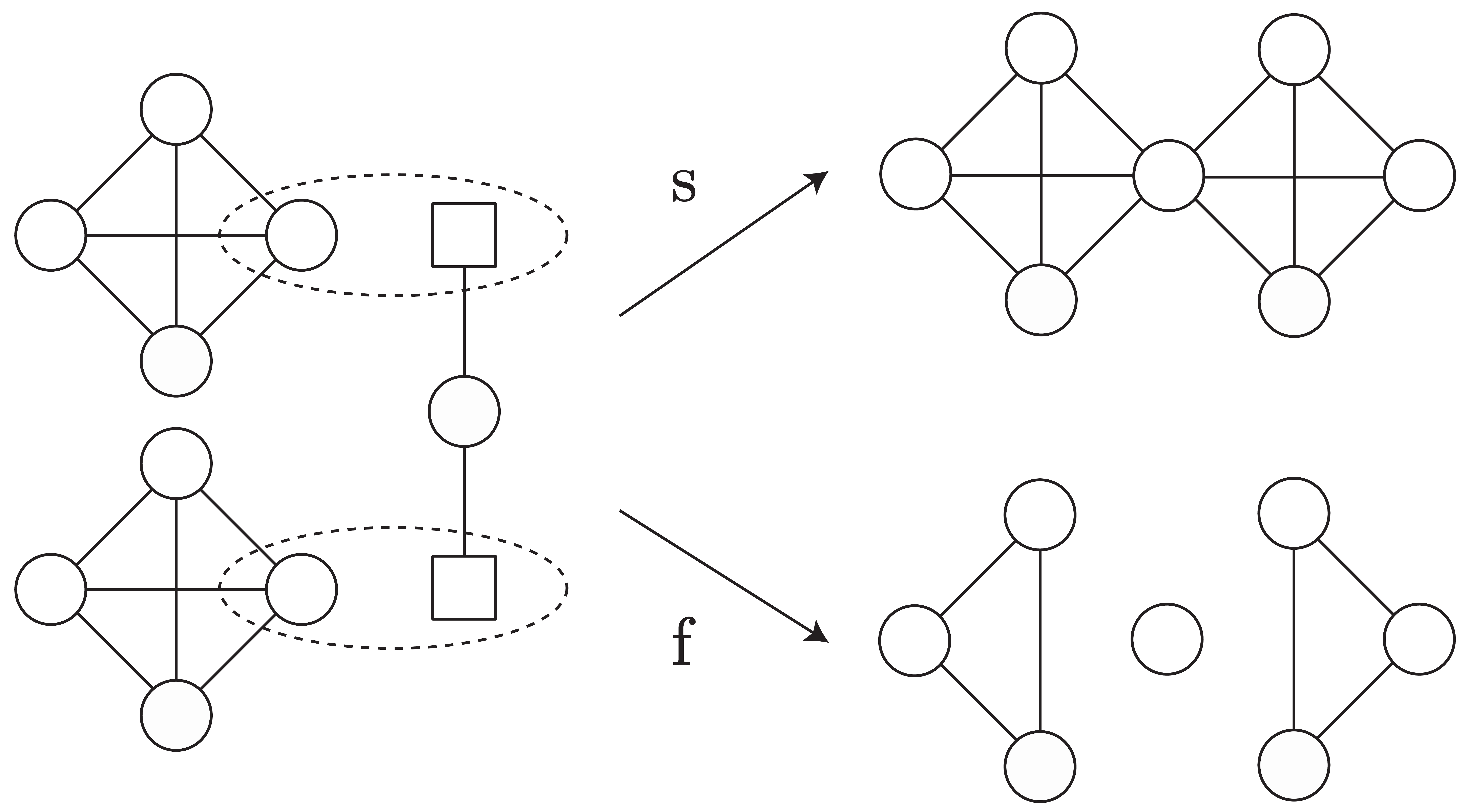}
\end{tabular}
\end{center}
\caption{The $3$-qubit chain is used as an intermediary to fuse two tetrahedra together to form a ``bowtie". Since there are two Bell measurements involved, the theoretical success probability of this step is $56.25$\%.}
\label{fig:TetrahedraFusion2}
\end{figure}

 This step succeeds with a probability of $75$\%. Next, we take the output of the first step (e.g. two tetrahedra), and fuse them together using a $3$-qubit chain, as shown in Fig. [\ref{fig:TetrahedraFusion2}]. 
The $3$-qubit chain acts as an intermediary in fusing two tetrahedra and producing the correct geometry of the resulting lattice that we desire. Since a Bell measurement failure on a $3$-qubit chain with Hadamard applied to a target qubit results in the destruction of the entanglement between all the qubits of the chain, this fusion step succeeds only when both the Bell measurements succeed, which happens with a probability of $56$\%. 

\begin{figure}[htb]
\begin{center}
\begin{tabular}{c}
\includegraphics[width= \columnwidth]{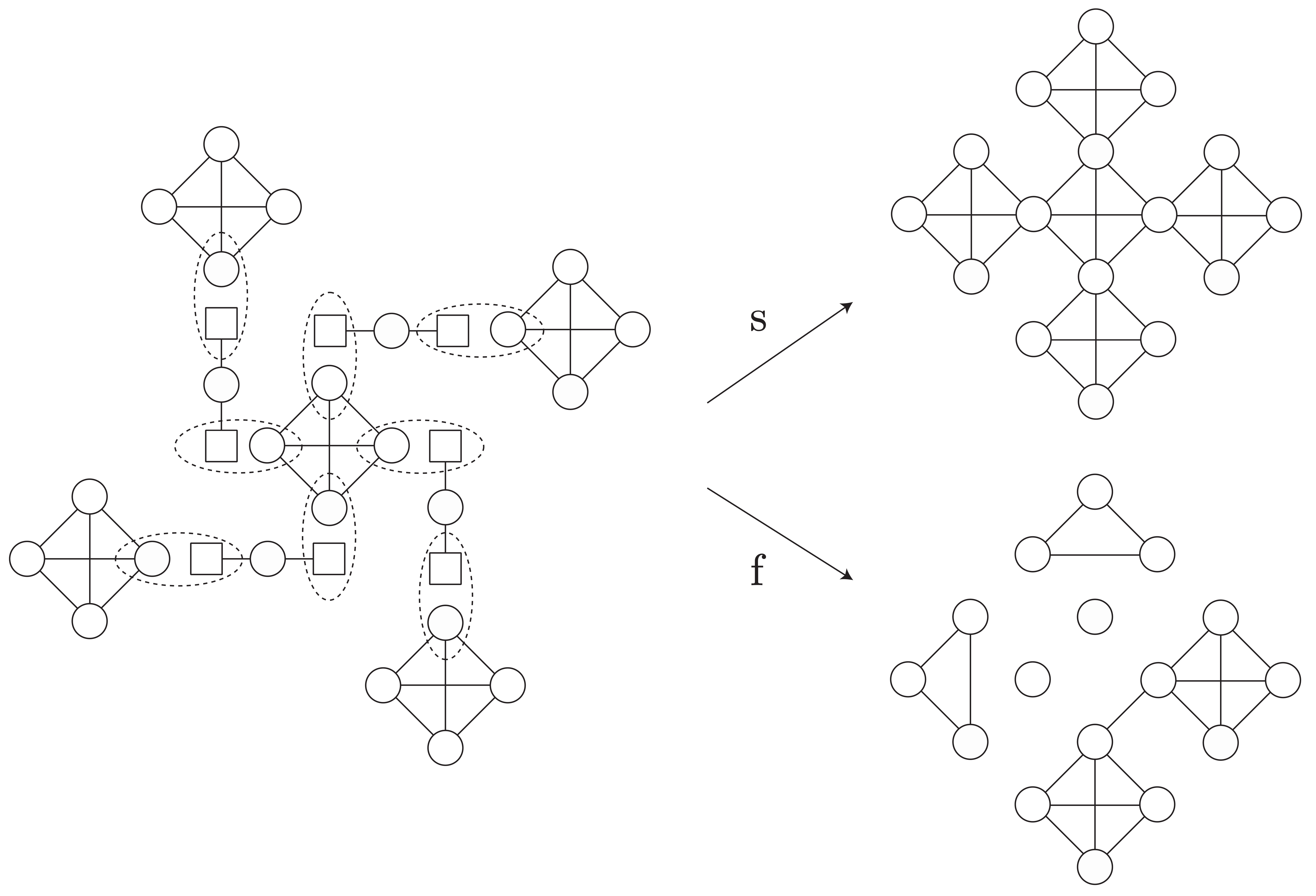}
\end{tabular}
\end{center}
\caption{$3$-qubit chains are used as intermediaries to fuse five tetrahedra together. The failure event is an example of what happens when the pair of fusions at the top and the top-left fail. For 3D views of a pyrochlore lattice, see Fig [\ref{fig:SimulationFigures} (a) and (b)].}
\label{fig:TetrahedraFusion3_1}
\end{figure}

The above two steps, outlined in Figs. [\ref{fig:TriangleFusion}] and [\ref{fig:TetrahedraFusion2}], can be repeated multiple times to create a full-fledged pyrochlore lattice of the desired size. Fig. [\ref{fig:TetrahedraFusion3_1}] outlines the repeated fusion process to create a larger pyrochlore lattice.

It is important to emphasize that the fusion steps proceed regardless of whether the previous steps fail or succeed. For example, if the fusion of a pair of triangular clusters failed, then we'd be fusing a Bell pair with, say, another tetrahedron using a $3$-qubit chain in Fig.[\ref{fig:TetrahedraFusion2}]. 
In that sense, we have not shown all the possible outcomes in the figures, though they can all be enumerated and described based on the discussion in the last section.

Refs. \cite{kieling2007,kieling2008} deal with the process of renormalization in depth. Here, we briefly state that renormalization of the pyrochlore lattice percolated in this manner can be accomplished by considering (polynomially many) hypothetical cubic blocks of some chosen size overlaid on the pyrochlore lattice. By measuring out suitable qubits in the Z basis we can ensure spanning paths between the desired faces of the block as well as renormalize the qubits in the blocks to single qubits. As the number of blocks grows there is an increasing probability that one or more will fail to be connected, but this can be dealt with by simply increasing the block sizes. Because such an increase in size improves the spanning probability exponentially, we can conclude that a polynomial sized computation is feasible with polynomial resources.
 
\begin{figure}
\centering
\mbox{\subfigure[\mbox{ }Fully connected pyrochlore cell]{\includegraphics[width=1.5in]{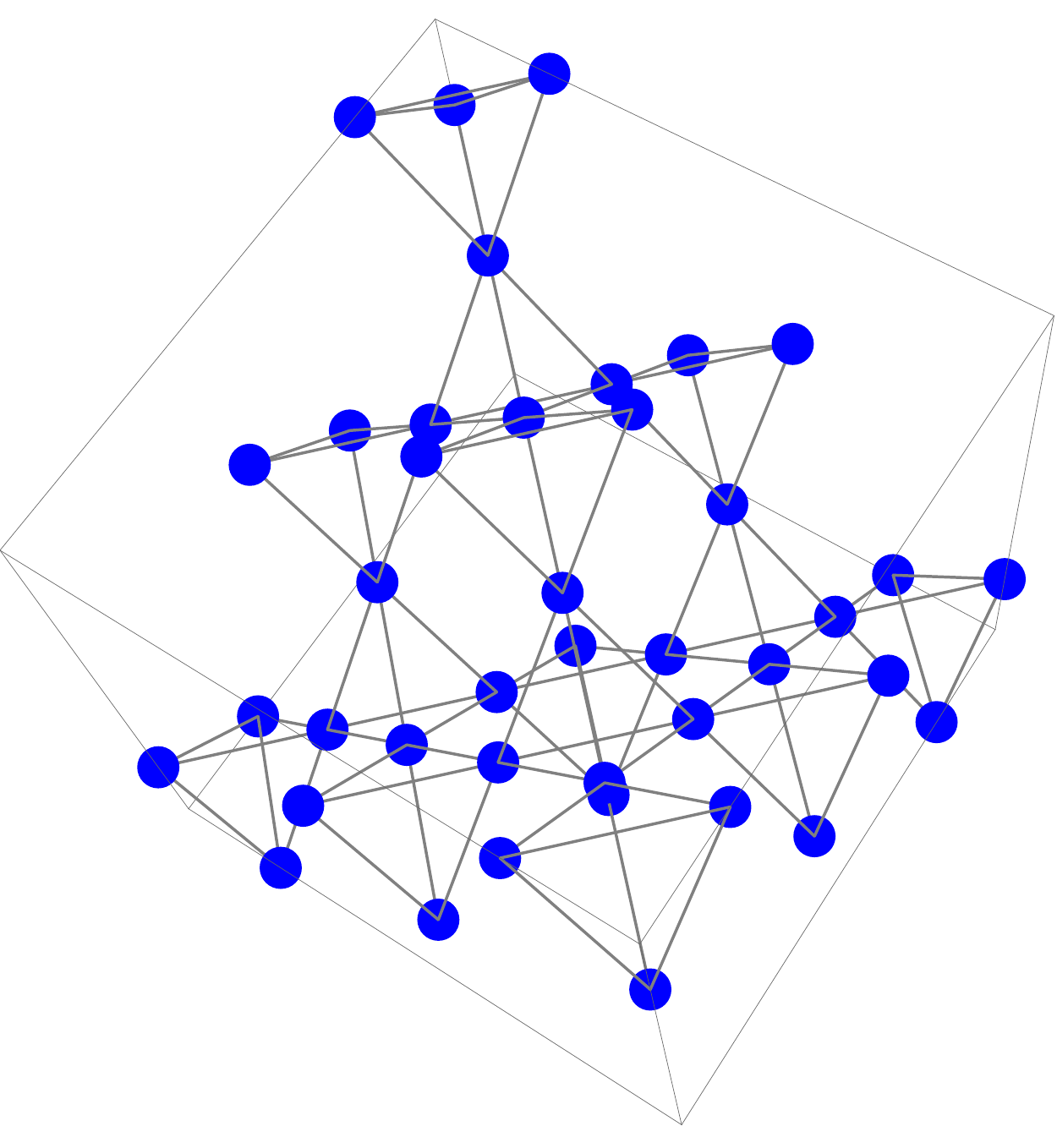}}\quad
\subfigure[\mbox{ }Another view of the cell in (a)]{\includegraphics[width=1.5in]{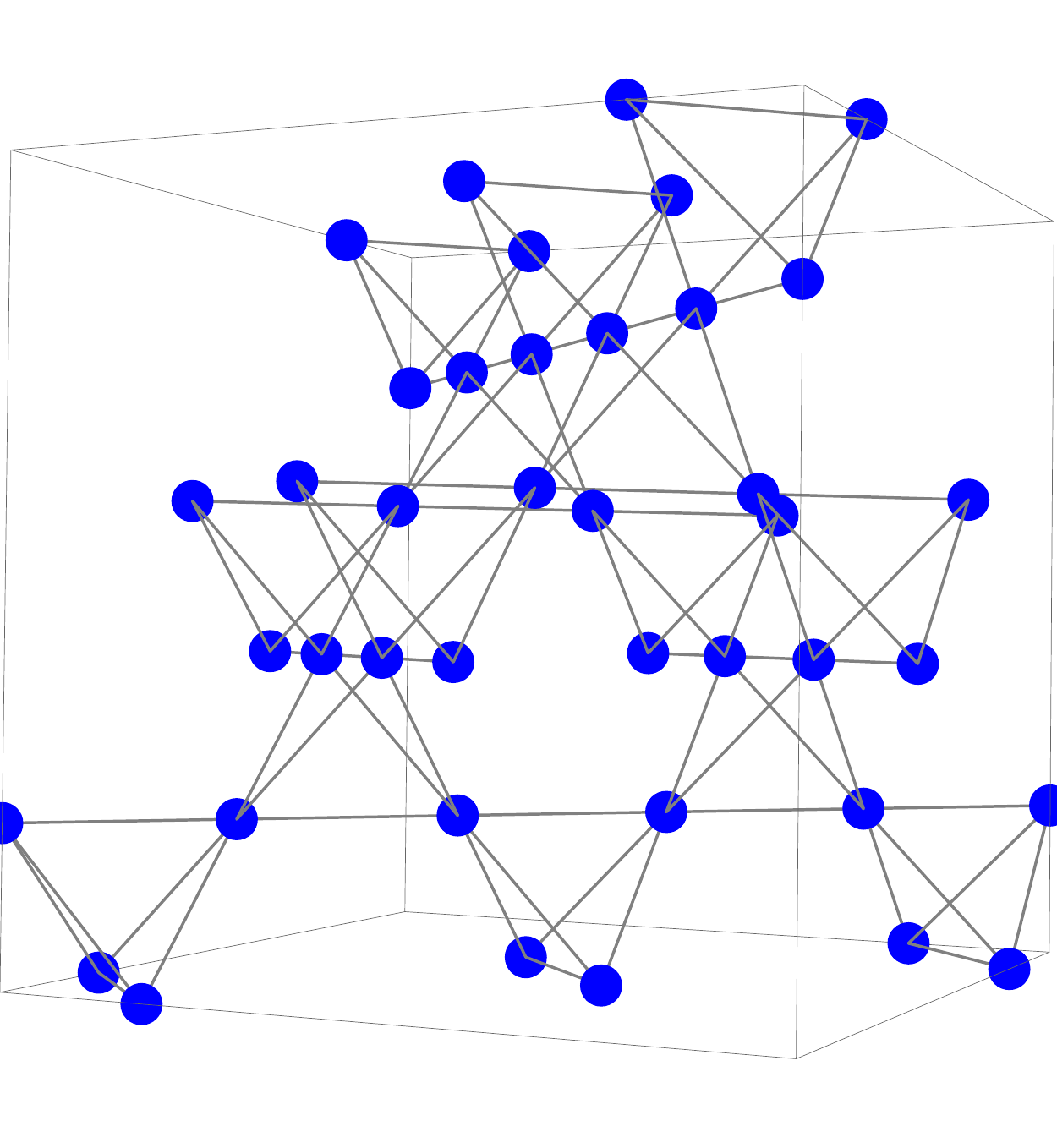} }\quad}\\
\mbox{\subfigure[\mbox{ }Effect of probabilistic creation of tetrahedra]{\includegraphics[width=1.5in]{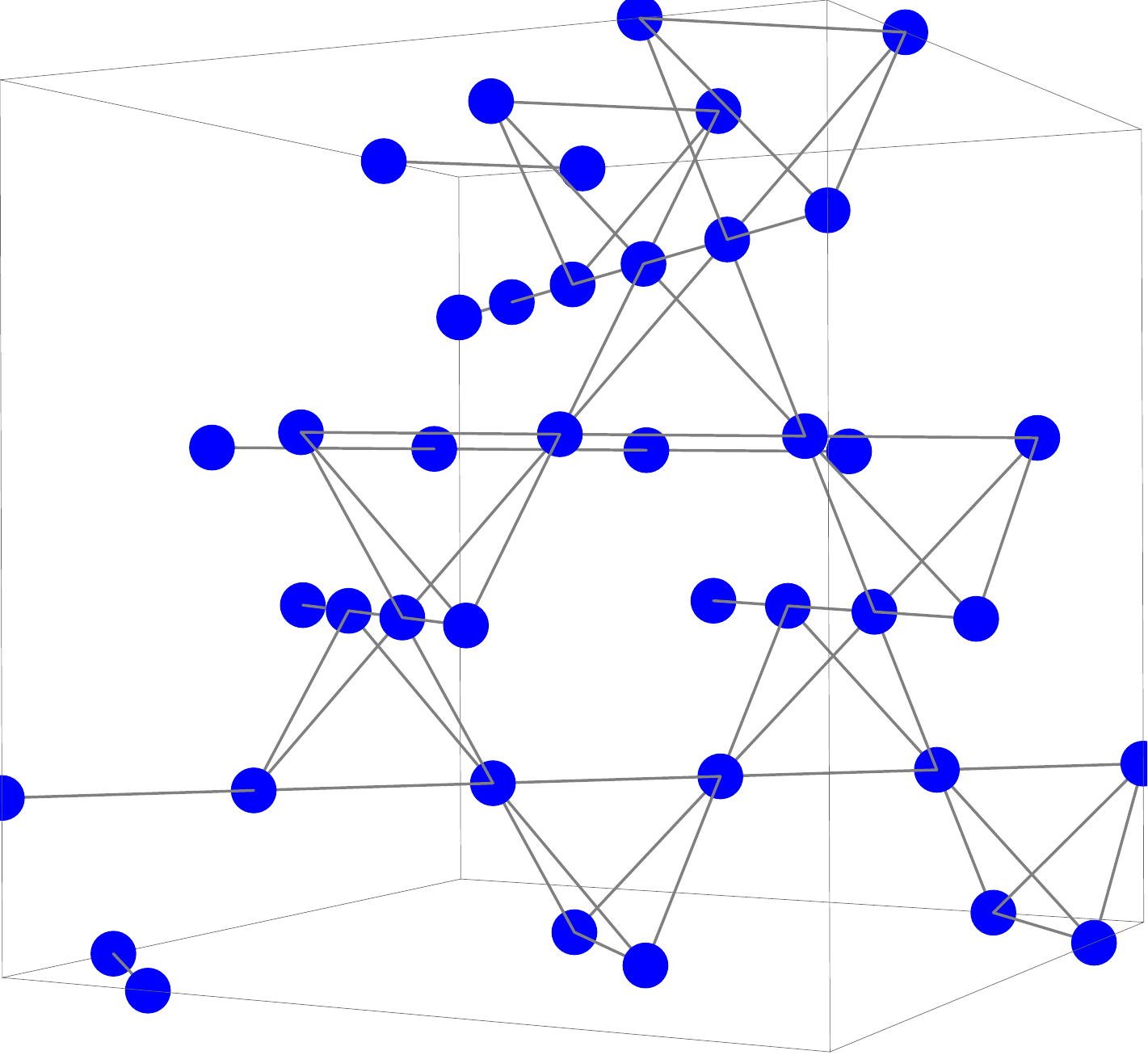} }\quad
\subfigure[\mbox{ }The lattice after probabilistic creation of tetrahedra and probabilistic fusion]{\includegraphics[width=1.5in]{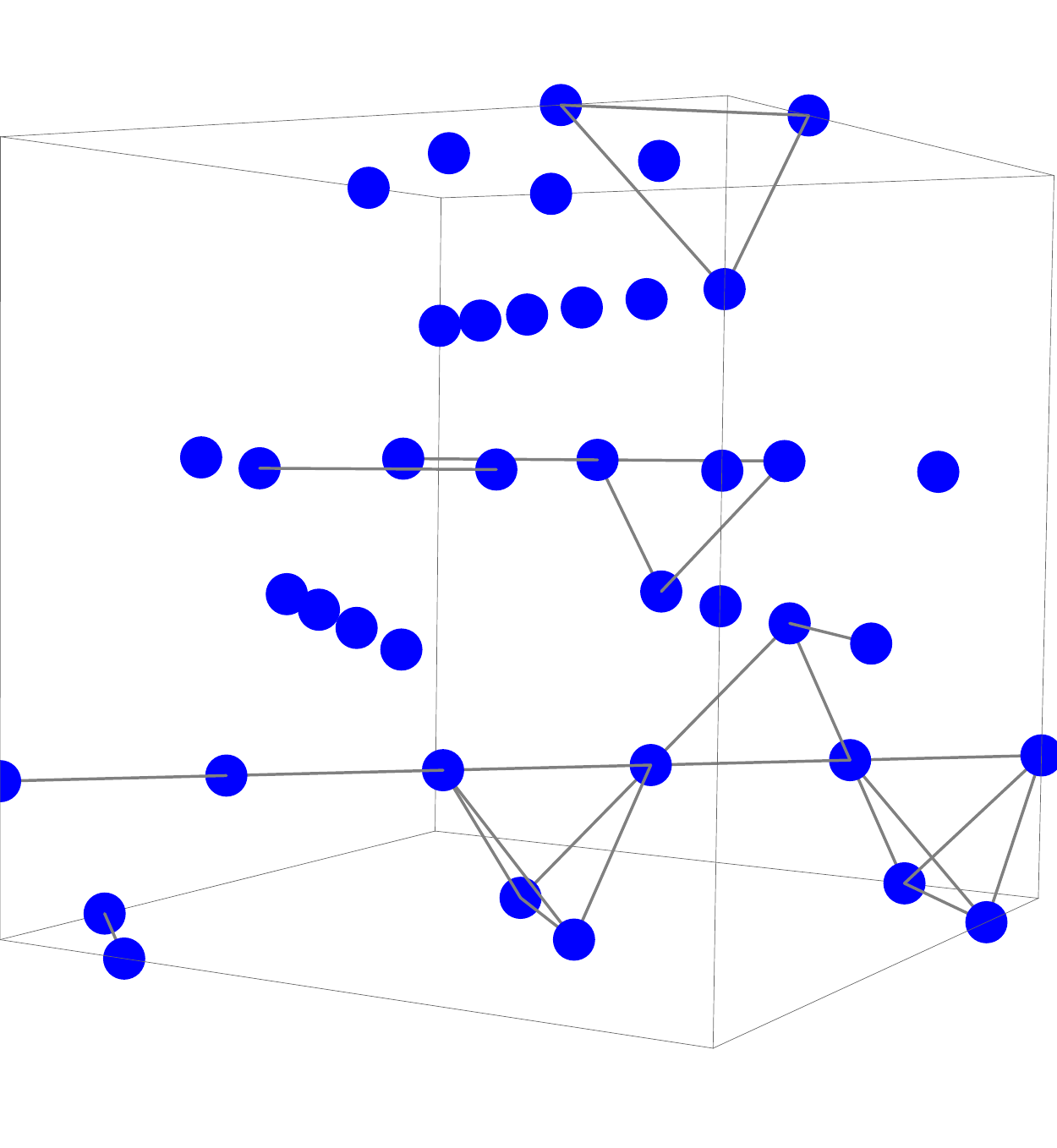} }\quad}
\caption{Simulation of our proposal of generating a spanning cluster using $3$-qubit chains and triangles. Figure (d) shows a representative outcome on a unit cell, while figure (a) is what we would produce if every step of our proposal was deterministic.}
\label{fig:SimulationFigures}
\end{figure}

\emph{Numerical Simulation---} 
Given that the above process cannot be mapped on to a simple vertex percolation or edge percolation problem, no analytical method for analyzing the percolation threshold for our scheme is known. Hence, we resort to simulating our scheme numerically in C++ \cite{dawson2014}. First, a fully connected $n_x\times n_y \times n_z$ pyrochlore lattice was built by translating the cell shown in Fig.[\ref{fig:SimulationFigures}(a),(b)] $n_x$, $n_y$, and $n_z$ times in the $x$, $y$ and $z$ directions, respectively. Hence, adding a unit cell corresponds to adding $26$ qubits to our simulation. Next, $25$\% of the tetrahedra were reduced to two disconnected ``lines", simulating the failure outcome of Fig.[\ref{fig:TriangleFusion}]. Finally, $56$\% of the vertices were disconnected simulating the failure outcome of Fig. [\ref{fig:TetrahedraFusion2}]. Fig. [\ref{fig:SimulationFigures} (c),(d)] show the effect of the two steps on a $1\times 1\times 1$ pyrochlore lattice (i.e. a pyrochlore ``unit cell" for our purposes).

\begin{figure}[htb]
\begin{center}
\begin{tabular}{c}
\includegraphics[width= 0.8 \columnwidth]{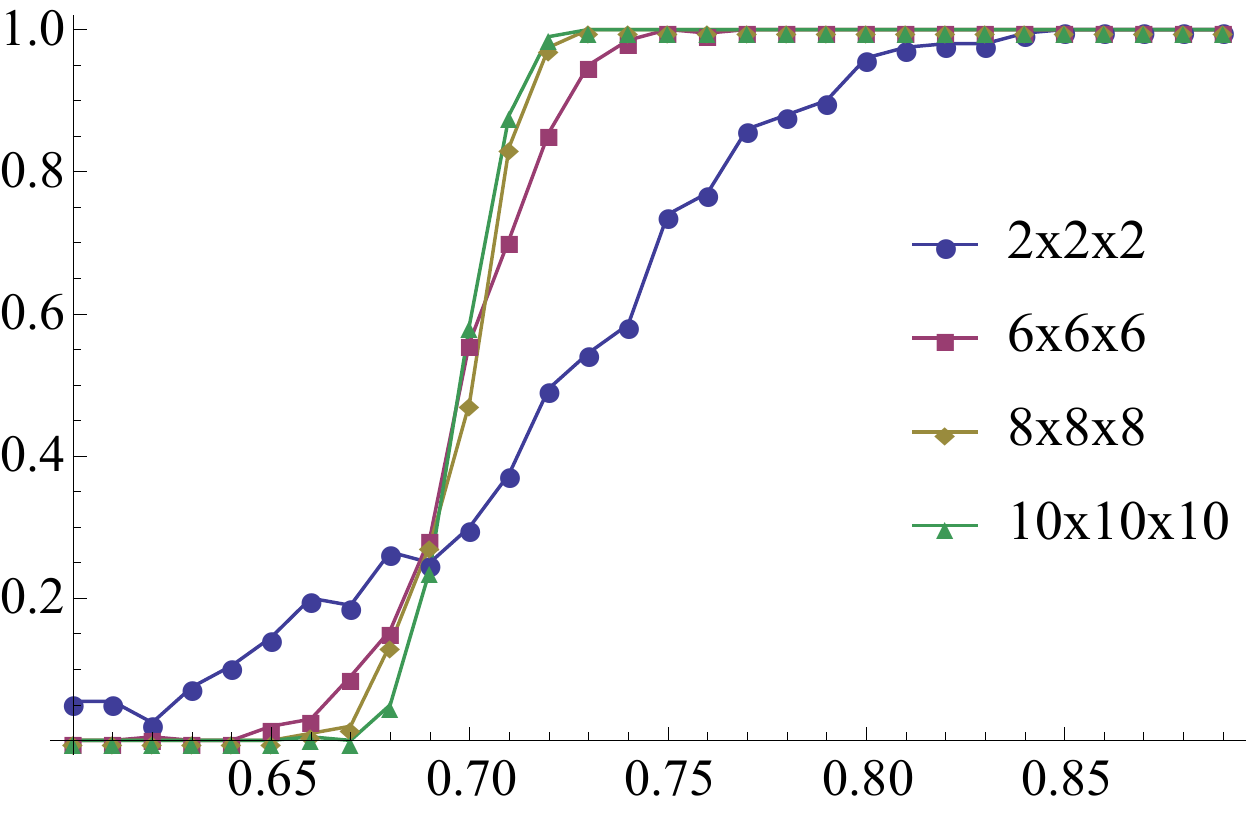}
\end{tabular}
\end{center}
\caption{Percolation probability versus the success probability of Bell measurement for various pyrochlore lattice sizes. Each lattice size was sampled 200 times.}
\label{fig:PyrochlorePercolationThresholdPlot}
\end{figure}

As seen in Fig. [\ref{fig:PyrochlorePercolationThresholdPlot}], the probability of having a spanning cluster in the x-direction approaches unity as the success probability of Bell measurement is increased from $65\%$ to $90\%$ (for a sample size of $200$ trials) and illustrates that the fusion probability of $75$\% is above the percolation threshold. 

\begin{figure}[htb]
\begin{center}
\begin{tabular}{c}
\includegraphics[width= 0.8 \columnwidth]{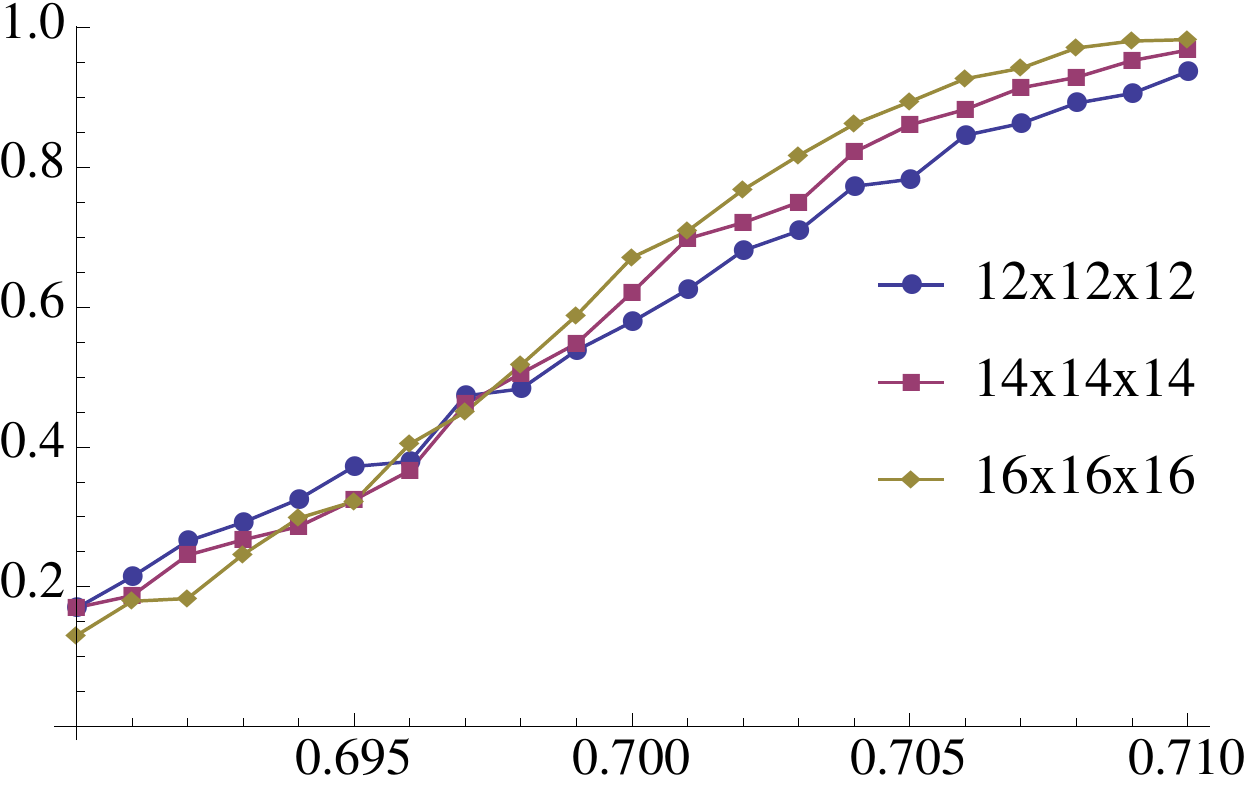}
\end{tabular}
\end{center}
\caption{Percolation probability versus the success probability of Bell measurement for large pyrochlore lattices. Large number of trials (1000), large lattice sizes, and a small step size in the fusion probability (0.001) allow us to pin-point the percolation threshold.}
\label{fig:PyrochlorePercolationThresholdPlotFineScale}
\end{figure}

To accurately estimate the percolation threshold, we simulated our protocol on lattice sizes of $12\times 12\times 12$, $14\times 14\times 14$, and $16\times 16\times 16$, the results of which are shown in Fig. [\ref{fig:PyrochlorePercolationThresholdPlotFineScale}]. The figure shows that the percolation threshold is between fusion probabilities of $69.5\%$ $70.0\%$, which is lower than the $75\%$ theoretical fusion success probability we have assumed for our protocol.

From an experimental point of view, we would like to know how big the underlying lattice needs to be in the $y$ and $z$ directions for a given length in $x$ so that the probability of having a spanning cluster in the $x$ direction is above, say, $90$\%. We also want to know the average size of a spanning cluster for a given $n_x$, $n_y$ and $n_z$ to know the scaling of resources. To answer these questions, we ran numerical simulations for various values of $n_x$, $n_y$ and $n_z$. The results of these simulations are shown in Table [\ref{tab:Scaling}]. As mentioned before, adding a unit cell corresponds to adding $26$ qubits, because of which our scaling results give us a course resolution in the scaling of the lattice size. Regardless, the simulations show that the size of the lattice can be kept to a few unit cells in the $z$ direction (roughly 3) to be above the $90$\% threshold for any required spanning length in the $x$ direction if the size in $y$ is of the order of the size in $x$. That is important for potential experiments since in optical-chip experiments it is desirable to keep the geometry close to planar \cite{matthews2009}.

\begin{table}[h]
\caption{Scaling of lattice size in $y$ and $z$ as a function of the length in $x$. We set $90$\% as the minimum probability threshold for having a spanning path. The success probality was calculated from $200$ samples for each value of $n_x$, $n_y$ and $n_z$} 
\centering 
\setlength{\tabcolsep}{8pt}
\begin{tabular}{c c c c c} 
\hline\hline 
       &       &       & Lattice & Success\\
$n_x$ & $n_y$ & $n_z$ & size & Probability (\%)
\\[0.5ex]
\hline 
4 & 4 & 4 & 1444 & 96\\
5 & 5 & 3 & 1680 & 91\\
6 & 6 & 3 & 2352 & 91\\
7 & 7 & 3 & 3136 & 92\\
8 & 7 & 3 & 3556 & 93\\
9 & 7 & 3 & 3976 & 90\\
10 & 8 & 3 & 4984 & 94\\
11 & 8 & 3 & 5460 & 94\\
12 & 9 & 3 & 6636 & 96\\
13 & 9 & 3 & 7168 & 92\\
14 & 9 & 3 & 7700 & 92\\
15 & 9 & 3 & 8232 & 91\\[0.5ex]
\hline 
\end{tabular}
\label{tab:Scaling}
\end{table}

\emph{Discussion and Future Directions---} We have proposed  creating a percolated pyrochlore cluster using Bell measurements with $75$\% success probability. This is a significant improvement over earlier proposals because of two reasons. First, we have brought down the initial resources required from $4$-qubit to $3$-qubit cluster state, which is the minimal required size for generating universal cluster states using Bell measurements. And second, Bell measurements are more robust to errors than Type-I fusions that were used in earlier proposals. The question of the quantitative effects of experimental errors on our protocol, and how these errors can be corrected, still needs to be answered. In photonics the two largest errors are systematic imperfections (interferometers that are misaligned) and photon losses, plus some small amount of Pauli error from any active elements (primarily switches) that a photon has to pass through \cite{matthews2009, berry2009}. The repeatability of systematic errors makes them relatively benign, and the fact that losses are always ultimately detected means that correcting for them is considerably easier than for unknown Pauli error. As such, ballistic protocols such as ours which limit the amount of active switching are preferable, however finding a way to combat all sources of error simultaneously in this architecture is ongoing work \cite{varnava2008}.

\emph{Appendix---}
Pick a qubit $i$ that neighbours qubit $1$ and qubit $j$ that neighbours qubit $1^\prime$ as shown in  Fig. [\ref{fig:BellMeasurementResult}]. We apply a Bell measurement on qubits $1$ and $1^\prime$. Qubit $1$ could have multiple neighbours in addition to $i$, and similarly qubit $1^\prime$ could neighbour multiple qubits in addition to $j$. The stabilizers \cite{browne2006} for the qubits ($S$) can be written as 
\begin{subequations}
\begin{align}
&S_i= AX_iZ_1 &\text{  ,  }\qquad & S_1= BX_1Z_i\\
&S_j= Z_{1^\prime} X_jD &\text{  ,  }\qquad &S_{1^\prime}= X_{1^\prime}Z_jC,
\end{align}
\end{subequations}
where $A$, $B$, $C$, and $D$ signify products of Pauli $Z$ operators for neighbours of $i$, $1$, $1^\prime$, and $j$, respectively. Applying the hadamard on $1^\prime$ results in
\begin{subequations}
\begin{align}
&S_i= AX_iZ_1 &\text{  ,  }\qquad & S_1= BX_1Z_i\\
&S_j= X_{1^\prime} X_jD &\text{  ,  }\qquad &S_{1^\prime}= Z_{1^\prime}Z_jC.
\end{align}
\end{subequations}

Performing a Bell measurement on qubits $1$ and $1^\prime$ means that our projectors are, for instance, $X_1X_{1^\prime}$ and $Z_1Z_{1^\prime}$. For the measured state, the two new stabilizers that commute with the projectors are
\begin{subequations}
\begin{align}
&\bar{S}_i= S_iS_{1^\prime} = AX_iZ_jC \\
&\bar{S}_j= S_iS_j = BZ_i X_jD.
\end{align}
\end{subequations}
This means that there is a cluster-like bond between the qubits $i$ and $j$. The qubit $i$ inherits all its neighbours alongwith the neighbours of $1^\prime$, $j$ inherits all its neighbours along with the neighbours of $1$. This analysis holds for all qubits $i$ and $j$ that are neighbors of $1$ and $1^\prime$, respectively. It also holds for other projectors in the Bell basis up to local unitary operations. 

\emph{Acknowledgements---} H.Z. was supported by US AFOSR Grant No. FA9550-11-1-0297. He is thankful to the Institute of Physics at the University of Mainz for their hospitality. PVL was supported by QuOReP (BMBF) and HIPERCOM (ERA-Net CHIST-ERA). TR supported by the by the Vienna Science and Technology Fund (WWTF, grant ICT12-041) and the Army Research Office (ARO) grant No. W911NF-14-1-0133. The authors are thankful to Mercedes-Gimeno Segovia for helpful discussions.


\bibliography{UBSDSqueezingPRL}
\end{document}